\documentclass[letter]{aa}
\usepackage{graphicx}
\usepackage{txfonts}

\begin{document}

\title{The magnetic field and the location of the TeV emitter in Cygnus~X-1 and LS~5039}

\author{V. Bosch-Ramon\inst{1} \and D. Khangulyan\inst{1} \and F.~A. Aharonian\inst{2,1}} 

\institute{Max Planck Institut f\"ur Kernphysik, Saupfercheckweg 1, Heidelberg 69117, Germany; 
vbosch@mpi-hd.mpg.de; Dmitry.Khangulyan@mpi-hd.mpg.de \and
Dublin Institute for Advanced Studies, Dublin, Ireland; Felix.Aharonian@dias.ie}

\offprints{ \\ \email{vbosch@mpi-hd.mpg.de}}

\abstract
{Cygnus~X-1 and LS~5039 are two X-ray binaries observed at TeV energies. 
Both sources are compact systems, contain jet-like (radio) structures, 
and harbor very luminous O stars. A TeV signal has been found around 
the superior conjunction of the compact object in both objects, when the highest gamma-ray opacities are expected.}
{We investigate the implications of finding TeV emission from Cygnus~X-1 and LS~5039 
around the superior conjunction, since 
this can give information on the system magnetic field and the location of the TeV emitter.}
{Using the very high-energy spectra and fluxes 
observed around the superior conjunction in Cygnus~X-1 and LS~5039, we compute
the absorbed luminosity that is caused by pair creation in the stellar photon field for different
emitter positions with respect to the star and the observer line of sight.
The role of the magnetic field and electromagnetic cascading are discussed. 
For the case of inefficient electromagnetic cascading, the expected 
secondary synchrotron fluxes are compared with the observed ones at X-ray energies.}
{We find that, in Cygnus~X-1 and LS~5039, either the magnetic field 
in the star surroundings is much smaller than the one 
expected for O stars or the TeV emitter is located at a distance $> 10^{12}$~cm from the compact object.}
{Our results strongly suggest that the TeV emitters in Cygnus~X-1 and LS~5039 are located at the borders of the binary 
system and well above the orbital plane. This would not agree with those models for which the emitter 
is well inside the system, like the innermost-jet region (Cygnus~X-1 and LS~5039; microquasar scenario), 
or the region between the pulsar and the primary star (LS~5039; {\it standard}{} pulsar scenario).}

\maketitle

\section{Introduction}

Cygnus~X-1 and LS~5039 are high-mass X-ray binaries presenting radiation in the very high-energy (VHE) range (Aharonian et
al. \cite{aharonian05}; Albert et al. \cite{albert07}). Both sources show extended outflows that generate non-thermal radio
emission (Stirling et al. \cite{stirling01}; Paredes et al. \cite{paredes00}) and are located at distances of $\approx
2.1$~kpc (Ziolkowski \cite{ziolkowski05}) and  $\approx 2.5$~kpc (Casares et al. \cite{casares05}; C05 hereafter),
respectively. The primary object in these two systems is an O star  (Gies \& Bolton \cite{gies86}; C05), and the compact
object is a $\sim 20$~M$_{\odot}$ black hole in the case of Cygnus~X-1 (Ziolkowski \cite{ziolkowski05}) and is still of
unknown nature in the case of LS~5039 (C05). The detection of jet-like radio emitting structures has  led to the
classification of these sources as microquasars (Stirling et al. \cite{stirling01};  Paredes et al. \cite{paredes00}). Later
on, however, not detecting accretion features in the X-ray spectrum of LS~5039 has been interpreted as a hint of a
non-accreting pulsar (e.g. Martocchia et al. \cite{martocchia05}). On the other hand, Cygnus~X-1 is a firmly established
accreting source (e.g. Pringle \& Rees \cite{pringle72}). It is worth noting that LS~5039 has been associated to an EGRET
source (Paredes et~al. \cite{paredes00}), whereas Cygnus~X-1 has no GeV association.

Cygnus~X-1 and LS~5039 are relatively similar to each other, as shown in Table~1. Besides showing extended radio emission, the
two systems harbor massive and hot stars, have quite compact size, $\sim 0.1$--0.2~AU, and have shown a TeV signal around the
superior conjunction of the compact object (SUPC). The TeV emission properties of both sources around SUPC are presented as
well in Table~1 (bottom). In the case of Cygnus~X-1, evidence of detection above 100~GeV of 4.1~$\sigma$ significance has
been reported once right before SUPC, at phase $\approx 0.9$ (Albert et al. \cite{albert07}). In LS~5039, the VHE emission
has been clearly detected (40~$\sigma$) all along the orbit, being in fact periodic (Aharonian et al. \cite{aharonian06})
with the orbital period (C05). At the phase $0.00\pm 0.05$ ($\sim$ SUPC), the detection significance is 6.1~$\sigma$. There
is however a big difference between the two sources. Whereas Cygnus~X-1 shows a thermal (comptonized) hard X-ray spectrum
(e.g. Sunyaev \& Trumper \cite{sunyaev79}), with a luminosity $L_{\rm X}\sim 10^{37}$~erg~s$^{-1}$ in the epoch when the TeV
emission was detected (Albert et al. \cite{albert07}), the X-ray radiation from LS~5039 is three orders of magnitude less
luminous, i.e. $L_{\rm X}\sim 10^{34}$~erg~s$^{-1}$, and could be dominated by a non-thermal component (e.g. Bosch-Ramon
et~al. \cite{bosch07}). 

In Cygnus~X-1 and LS~5039, if the TeV emitter is deep inside the system and behind the primary star around SUPC, the
photon-photon absorption opacity ($\tau$) for TeV gamma rays in the stellar photon field will be $\gg 1$ in the direction  of
the observer. In this situation, either the magnetic field is low enough to allow efficient electromagnetic (EM) cascading to
develop (e.g. Aharonian et al. \cite{aharonian06b}; Bednarek \& Giovanelli \cite{bednarek07}; Orellana et al.
\cite{orellana07}), effectively reducing $\tau$ around SUPC, or the absorbed energy will be reradiated mainly by secondary
pairs via synchrotron emission (e.g. Bosch-Ramon, Khangulyan, \& Aharonian \cite{bosch08}). In the latter case, the high
(effective) opacities will require an injected gamma-ray luminosity much higher than the observed one, and the
synchrotron luminosity of the secondary pairs will be similar to that of the absorbed gamma rays. Because the secondary
synchrotron luminosity should not exceed the observational constraints, we can put strong restrictions on either the magnetic
field strength or on the emitter location with respect to the star and the line of sight of the observer. In this work, we
discuss the implications of two possible situations, i.e. a {\it low} versus a {\it high} magnetic field in the stellar
surroundings for the TeV emitters in Cygnus~X-1 and LS~5039. 

\section{The magnetic field in the stellar surroundings}\label{abs}

If the TeV emitter moves along the orbit {\it close} to the compact object, the maximum in photon-photon absorption will take
place around SUPC (e.g. see Fig.~14 in Khangulyan, Aharonian \& Bosch-Ramon \cite{khangulyan08} -K08 hereafter-). For sources
like Cygnus~X-1 and LS~5039, $\tau$ is $\approx 20$ for $i=45^{\circ}$ and 300~GeV photons emitted in the vicinity of the
compact object, around SUPC. This shows that approximately $10^9$ times more radiation should be produced to obtain the
observed gamma-ray fluxes. This estimate gives an idea of the amount of energy that can be required to power the observed VHE
radiation. 

The required energy budget to power the VHE emission around SUPC can be significantly reduced through efficient EM cascading,
which occurs if IC scattering takes place deep enough in the Klein Nishina (KN) regime, as is the case for electrons of
energy $\ga$~TeV in the stellar photon field.  Nevertheless, the dominance of KN IC energy losses for TeV electrons requires
a magnetic field strength well below a critical value, $B_{\rm c}$,  since otherwise a substantial fraction of the energy
will be radiated via the synchrotron process. The value of 
$B_{\rm c}$ for TeV electrons is defined by the balance of synchrotron and IC
energy losses (K08):  
\begin{equation} 
B_{\rm c}\approx
10\,\left(\frac{L_*}{10^{39}~{\rm erg~s}^{-1}}\right)^{1/2}\,\left(\frac{R_*}{R}\right)\,{\rm G}\,,
\end{equation} 
where $R$, $R_*$, and $L_*$ are the star distance, radius, and luminosity, respectively. 

To know whether efficient EM cascading can occur requires the magnetic field strength expected in the surroundings of the TeV
emitter, $B_*$, at $1-2\,R_*$ from the O-star surface. The magnetic fields close to the stellar surface has been only
directly measured in several O stars ($\theta^1$~Ori~C, Donati et al. \cite{donati06}; HD~191612, Wade et al. \cite{wade06};
and $\zeta$~Orionis~A, Bouret et al. \cite{bouret08} for clear detections; and HD~36879, HD~148937, HD~152408, and H~D164794,
Hubrig et al. \cite{hubrig08} for 3--4~$\sigma$ detections) due to the difficulties detecting the Zeeman effect, finding
however relatively high values of $\sim 100$--1000~G. In addition, it has been speculated that magnetic fields could be a
common feature of massive stars, as inferred from X-ray photometric and spectroscopic results (e.g. Stelzer et al.
\cite{stelzer05}; Waldron \& Cassinelli \cite{waldron07}), non thermal radio synchrotron emission (e.g. Benaglia
\cite{benaglia05}; Schnerr et al. \cite{schnerr07}), cyclical variations in UV wind spectral lines (e.g. Fullerton
\cite{fullerton03}; Kaper et al. \cite{kaper96}), or magnetic fields in neutron stars assuming a fossil origin (e.g. Ferrario
\& Wickramasinghe \cite{ferrario06}). Finally, the $R$-dependence of $B_*$ could be, roughly, $\propto 1/R^{1,2,3}$ depending
on $R$ (i.e. a toroidal, radial, or poloidal dominant component; Usov \& Melrose \cite{usov92}). Even in the extreme case of
$B_*\propto 1/R^3$ and 100~G at the stellar surface, $B_*\ga B_{\rm c}$ at $R\sim 2\,R_*$, which would prevent efficient EM
cascading (although $B_*\ll B_{\rm c}$ still cannot be discarded).

\section{Location of the emitter}

As seen in Sect.~\ref{abs}, under $B_*>B_{\rm c}$ most of the absorbed energy is reemitted via synchrotron emission. The
corresponding characteristic frequency and the short radiation timescales of secondary pairs with energies around or a bit
larger than the pair creation threshold ($\sim m_{\rm e}^2c^4/3\,kT_*$), 100--1000~GeV, determine that the synchrotron
emission will peak at X-rays or will flatten in the X-ray to gamma-ray range (K08; Bosch-Ramon et al. \cite{bosch08}). This
cannot be prevented by particle escape, since the radiative timescales at the relevant particle energies will be shorter than
the escape ones even when particles move at $c$. That this X-ray secondary synchrotron emission cannot overcome the
observed X-ray fluxes restricts the location of the TeV emitter.

\begin{figure}[]
\begin{center}
\resizebox{\hsize}{!}{\includegraphics{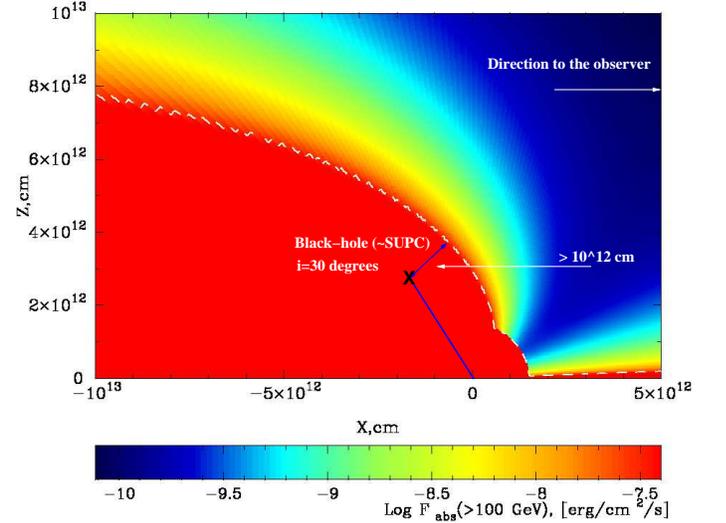}}
\caption{Two-dimensional map for Cygnus~X-1 of the amount of flux absorbed via photon-photon interactions 
for different TeV emitter locations within the system. The XY-plane is
the one that joins the observer, the star, and the emitter, 
and the X-direction is the one joining the star and the observer (to the right).  The {\bf X} in the plot
represents the location of the compact object at SUPC for an inclination angle of 30$^{\circ}$. The region to the left of the long-dashed line is forbidden 
by the constraints from X-ray observations
as a location of the TeV emitter for $B_*$ well above $B_c$. The emitter has to be at a distance $>10^{12}$~cm between the compact 
object and the closer point in the X-ray flux limit curve.}
\label{cygmap}
\end{center}
\end{figure}

\begin{figure}[]
\begin{center}
\resizebox{\hsize}{!}{\includegraphics{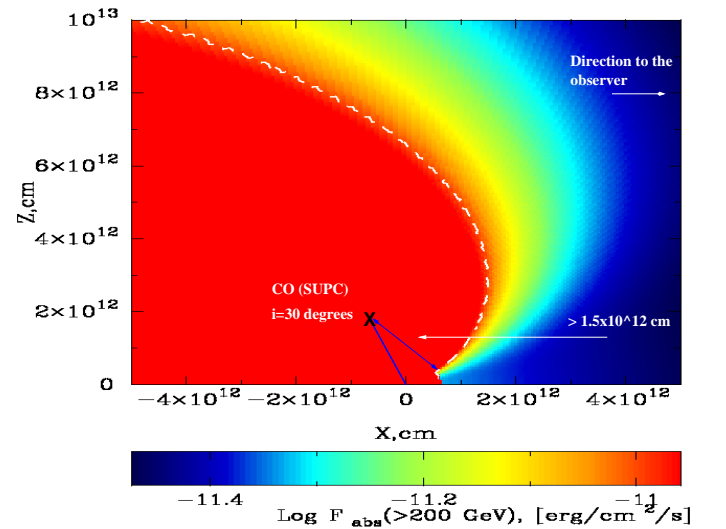}}
\caption{The same as in Fig.~\ref{cygmap} but for LS~5039. In this case, 
the emitter location is constrained to distances over a few $10^{12}$~cm.}
\label{lsmap}
\end{center}
\end{figure}

To estimate the amount of energy that could be released via synchrotron emission in X-rays, we plotted 2-dimensional maps of
the absorbed luminosity depending on the location of the emitter in the binary system. These maps are presented in
Figs.~\ref{cygmap} and \ref{lsmap} for Cygnus~X-1 and LS~5039, respectively. The XY coordinates correspond to the plane
formed by the emitter and star positions and by the observer line of sight. To compute these maps, we {\it deabsorbed}
the observed spectra and fluxes $\ga 100$~GeV around SUPC in both Cygnus~X-1 and LS~5039 (see Table~1). The regions forbidden
by the X-ray observational constraints in Cygnus~X-1 and LS~5039 are limited by a contour line. We computed the
2-dimensional maps for both sources assuming an anisotropic (e.g. stellar photon IC) and an isotropic mechanism for the
production of primary gamma rays. Both cases gave quite similar results, thus we chose the most conservative one for
each source, i.e. an anisotropic mechanism in the case of Cygnus~X-1 and an isotropic one in the case of LS~5039. We also
explored the impact of adopting the observational lower- and upper-limits for the spectral slopes (see Table~1), obtaining
only small differences between the resulting maps. Therefore, we adopted the mean observed 
spectral indexes to create the plots
presented in Figs.~\ref{cygmap} and \ref{lsmap}. 

The computed maps are $i$-independent, since they show the amount of absorbed luminosity for different emitter locations in
the emitter-star-observer plane, wherever the compact object is located. However, for illustrative purposes, we show the
position of the compact object in the emitter-star-observer plane when $i=30^{\circ}$. This plus the contour lines give an
idea of where the emitter could be with respect to the compact object. As seen in Figs.~\ref{cygmap} and \ref{lsmap}, the TeV
emitter can hardly be closer than $\approx 10^{12}$~cm from the black hole in Cygnus~X-1. For the case of LS~5039, the
minimum distance is a bit larger. It is remarkable that the energy requirements for an emitter close to the compact object
grow for larger $i$ (i.e. towards a neutron star mass for the compact object -C05-), since larger inclinations locate the
compact object deeper in the {\it forbidden} regions (see also Fig.~6 in B\"ottcher \cite{boettcher07}). We also note that,
since $\tau\propto 1/R$ (e.g. K08), the energy requirements would even be stronger if the TeV emitter were located in the
region between the compact object and the star.

\begin{figure}[]
\begin{center}
\resizebox{\hsize}{!}{\includegraphics{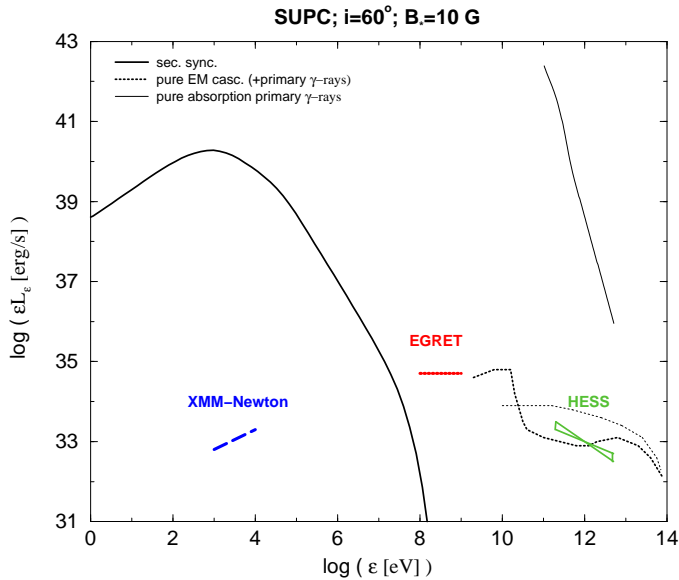}}
\caption{The SED of the 
synchrotron emission (solid line)
produced by the secondary pairs (derived from the {\it deabsorbed} primary gamma-ray spectrum) 
and the SED for pure EM cascading (dotted line) are
presented for LS~5039. 
The TeV emitter is located close to the compact object at SUPC, and the inclination angle is 
$i=60^{\circ}$ and, for the secondary emission case, $B_*=10$~G. 
The absorbed luminosity in the 
secondary case is about $\sim 10^{42}$~erg~s$^{-1}$. For the pure cascade case, the injected luminosity in 
primary gamma rays is $\sim 3\times 10^{35}$~erg~$s^{-1}$. 
The SEDs observed by {\it XMM-Newton} (X-rays) close to SUPC (at phase 0.0; Bosch-Ramon et al. \cite{bosch07}), EGRET (GeV)
(time averaged; Hartman et al. \cite{hartman99}), and HESS (phase interval: 0.0--0.1; Aharonian et al. \cite{aharonian06}), 
are also shown.}
\label{sed}
\end{center}
\end{figure}

In Fig.~\ref{sed}, we show the spectral energy distribution (SED) of the synchrotron radiation from the secondary pairs in
LS~5039 at SUPC, assuming that the TeV emitter is in the vicinity of the compact object. The {\it deabsorbed} and the primary
gamma-ray spectra are presented as well. We took $i=60^{\circ}$, which would correspond to the non-accreting pulsar
scenario, and $B_*=10$~G\footnote{The magnetic field can be taken as homogeneous in the secondary synchrotron emitter,
since the large $\tau$ implies that most of the radiation originates in a small region, of size $\ll R$.}. In the same plot,
for illustrative purposes, a computed pure IC cascade SED ($B_*\ll B_{\rm c}$), roughly similar to the observed VHE SED at
SUPC, is shown. As can be seen in the figure, pure absorption and secondary emission renders very large synchrotron X-ray
fluxes, far above the observed ones. Otherwise, the occurrence of pure EM cascading can effectively reduce $\tau$ by several
orders of magnitude, and it yields as well GeV luminosities consistent with the EGRET ones (see also Aharonian et al.
\cite{aharonian06b}). In the case of Cygnus~X-1, the secondary synchrotron SED, not shown here, would look similar to
LS~5039, although the observed X-ray fluxes would appear $\sim 4$ orders of magnitude below the computed X-ray fluxes,
instead of the $\sim 8$ of LS~5039. 

We emphasize that the precise shape of the observed spectra, which is 
difficult to obtain from the low statistics of the data around SUPC, is not important for our calculations, provided
that the {\it deabsorbed} spectrum is very soft, as seen in Fig.~\ref{sed} for LS~5039. In the case of Cygnus~X-1 (not shown here), the {\it deabsorbed} SED is roughly similar to
the one of LS~5039.

 \begin{table}[]
  \begin{center}
  \caption[]{Properties of the binary systems Cygnus~X-1 and LS~5039}
  \label{tab2}
  \begin{tabular}{lll}
  \hline\noalign{\smallskip}
  \hline\noalign{\smallskip}
 & Cygnus~X-1 & LS~5039 \\
  \hline\noalign{\smallskip}
Star luminosity [erg~s$^{-1}$] & $1.3\times 10^{39}\,^b$ & $7\times 10^{38}\,^a$ \\
Star temperature [K] & $3\times 10^4\,^b$ & $3.8\times 10^4\,^a$ \\
Stellar radius [cm] & $1.5\times 10^{12}\,^b$ & $7\times 10^{11}\,^a$ \\
Orbital semi-major axis [cm] & $3.5\times 10^{12}\,^b$ & $2.1\times 10^{12}\,^a$ \\
Orbital distance at SUPC [$R_*$] & $\approx 2.3$ & $\approx 2$ \\
Distance [kpc] & $2.1^b$ & $2.5^a$ \\
Eccentricity & $0^c$ & $0.35^a$ \\
Inclination & $30^{\circ}\,^d$ & $\sim 20^{\circ}-60^{\circ}\,^a$ \\
Wind velocity [cm~s$^{-1}$] & $\approx 2\times 10^8\,^{e}$ & $\approx 2\times 10^8\,^a$ \\
Mass loss rate [M$_{\odot}$~yr$^{-1}$] & $\approx 2\times 10^{-6}\,^{f}$ & $\approx 5\times 10^{-7}\,^{a}$ \\
  \hline\noalign{\smallskip}
Luminosity [$>100$~GeV]; erg~s$^{-1}$] & $\approx 1.6\times 10^{34}\,^g$ & $\approx 4\times 10^{33}\,^h$ \\
Spectral index & $3.2\pm 0.6^g$ & $2.6\pm 0.3^h$\\
  \noalign{\smallskip}\hline
  \end{tabular}
  \end{center}
{
$^{a}$ C05
$^{b}$ Ziolkowski (\cite{ziolkowski05})
$^{c}$ Gies \& Bolton (\cite{gies82})
$^{d}$ Gies \& Bolton (\cite{gies86})
$^{e}$ Herrero et al. (\cite{herrero95})
$^{f}$ Gies et al. (\cite{gies03})
$^{g}$ Albert et al. (\cite{albert07})
$^{h}$ Aharonian et al. (\cite{aharonian06})
}  
\end{table}

\section{Discussion}

The detection of Cygnus~X-1 and LS~5039 around SUPC, plus X-ray data, give strong constraints on the 
magnetic field and the location of the TeV emitter in these systems. In
summary, if the TeV emitter is close to the compact object, the magnetic field in the stellar surroundings must be smaller 
than a few Gauss. These values will allow IC cascades to
develop, effectively reducing the photon-photon absorption opacities. However, this magnetic field seems rather low when compared with the values expected in O stars. For a more
realistic $B_*>B_{\rm c}$, EM cascading becomes inefficient, and the large amount of energy absorbed in the stellar photon field will be reemitted by the secondary pairs mainly
via the synchrotron process. For $B_*>B_{\rm c}$, this radiation will peak roughly in the X-ray band. In such a case, to avoid the violation of the observational constraints, the
TeV emitter must be located at distances $>10^{12}$~cm from the compact object in both Cygnus~X-1 and LS~5039. It is consistent 
with how, as noted by K08, extreme
acceleration rates are required to explain the highest energy photons from LS~5039 if the accelerator is deep inside the system. Remarkably, the low hydrogen column densities
inferred using X-ray data would also hint at an X-ray emitter far from the compact object (Bosch-Ramon et al. \cite{bosch07}). 

A TeV emitter far from the compact object requires a physical mechanism to transport energy to the borders of the system to
power the VHE emission. The source of this energy could be accretion in Cygnus~X-1, or either accretion (microquasar
scenario) or a powerful pulsar wind (non-accreting pulsar scenario) in LS~5039. The energy carrier could be
the jet seen in radio in Cygnus~X-1. In the case of LS~5039, energy could be transported by a jet (the jet-like structure
seen in radio) powered by radiatively inefficient accretion (e.g. Bogovalov \& Kelner \cite{bogovalov05}), since no accretion
X-ray features have been found. In the {\it standard} pulsar scenario for LS~5039 (e.g. Dubus \cite{dubus08};
Sierpowska-Bartosik \& Torres \cite{sierpowska08}), the TeV emitter would be located close to the pulsar or around the line
joining the pulsar and the star. This would imply that, for reasonable energy budgets, very little emission should be
expected due to absorption around SUPC (see Fig.~2 in Dubus \cite{dubus06} -lower panel-, Fig.~4 in Dubus et al.
\cite{dubus08}, and Fig.~2 in Sierpowska-Bartosik \& Torres \cite{sierpowska08}), unlike how 
it is observed. This does not agree with the {\it standard}
pulsar scenario for LS~5039, although in a more general case, a supersonic outflow produced in the star/pulsar wind
colliding region (see Bogovalov et al. \cite{bogovalov08}) may transport energy efficiently to the system borders and farther
out. 

To conclude, our results strongly suggest that the TeV emitters in Cygnus~X-1 and LS~5039 are located at the borders of the
binary system, well above the orbital plane. This would not be compatible with those models for which the emitter is well inside
the system, like the innermost-jet region (Cygnus~X-1 and LS~5039; microquasar scenario) or the region between the pulsar
and the primary star (LS~5039; {\it standard} pulsar scenario).

\begin{acknowledgements} 
The authors thank the anonymous referee for constructive comments.
V.B-R. gratefully acknowledges support from the Alexander von Humboldt Foundation.
V.B-R. acknowledges support by DGI of MEC under grant
AYA2007-68034-C03-01, as well as partial support by the European Regional Development Fund (ERDF/FEDER).
\end{acknowledgements}

{}

\end{document}